\documentclass[conference]{IEEEtran}
\IEEEoverridecommandlockouts
\usepackage{cite}
\usepackage{amsmath,amssymb,amsfonts}
\usepackage{algorithmic}
\usepackage{algorithm} 
\usepackage{graphicx}
\usepackage{textcomp}
\usepackage{wrapfig}
\usepackage{xcolor}
\usepackage{makecell} 
\usepackage{multirow}
\usepackage{diagbox}
\usepackage{fancyhdr}
\usepackage{caption}
\usepackage{tikz}

\usepackage[a4paper, total={184mm,239mm}]{geometry}
\def\BibTeX{{\rm B\kern-.05em{\sc i\kern-.025em b}\kern-.08em
    T\kern-.1667em\lower.7ex\hbox{E}\kern-.125emX}}
\begin{document}

\title{FLICKER: A Fine-Grained Contribution-Aware Accelerator for Real-Time 3D Gaussian Splatting\\
\thanks{*  Corresponding author.}
}

\author{
\IEEEauthorblockN{Wenhui Ou\textsuperscript{1}, Zhuoyu Wu\textsuperscript{2}, Yipu Zhang\textsuperscript{1}, Dongjun Wu\textsuperscript{1}, Freddy Ziyang Hong\textsuperscript{1}, C. Patrick Yue\textsuperscript{1,*}}
\IEEEauthorblockA{\textit{\textsuperscript{1}Department of Electronic and Computer Engineering, The Hong Kong University of Science and Technology}\\
\textit{\textsuperscript{2}School of IT, Monash University, Malaysia Campus}
}
}

\maketitle
\begin{tikzpicture}[remember picture,overlay]
\node[anchor=north] at ([yshift=-8pt]current page.north)
{This work has been accepted at IEEE/ACM Design, Automation \& Test in Europe Conference (DATE) 2026};
\end{tikzpicture}

\begin{abstract}
Recently, 3D Gaussian Splatting (3DGS) has become a mainstream rendering technique for its photorealistic quality and low latency. However, the need to process massive non-contributing Gaussian points makes it struggle on resource-limited edge computing platforms and limits its use in next-gen AR/VR devices. A contribution-based prior skipping strategy is effective in alleviating this inefficiency, but the associated contribution-testing workload becomes prohibitive when it is further applied to the edge. In this paper, we present FLICKER, a contribution-aware 3DGS accelerator that leverages a hardware--software co-design framework, including adaptive leader pixels, pixel-rectangle grouping, hierarchical Gaussian testing, and mixed-precision architecture, to achieve near-pixel-level, contribution-driven rendering with minimal overhead. 
Experimental results show that our design achieves up to $1.5\times$ speedup, $2.6\times$ energy efficiency improvement, and $14\%$ area reduction over a state-of-the-art accelerator. Meanwhile, it also achieves $19.8\times$ speedup and $26.7\times$ energy efficiency compared with a common edge GPU.
\end{abstract}

\begin{IEEEkeywords}
3DGS, Contribution-Aware, Accelerator
\end{IEEEkeywords}

\section{Introduction}
\label{Sec:Introduction}
Advances in photorealistic novel view synthesis (NVS) have significantly enhanced the immersive experience in augmented and virtual reality (AR/VR) applications \cite{meta2024orion}. Recently, 3D Gaussian Splatting (3DGS) \cite{kerbl20233d,wu2024recent} has emerged as a leading NVS technique for its outstanding rendering speed. This makes it a promising solution for next-generation AR/VR systems. However, such real-time capability is largely limited to powerful cloud or desktop-level GPUs, while edge devices still suffer from severe performance loss and energy constraints \cite{lin2025metasapiens}.

The limitation stems from its inherent principle. Specifically, 3DGS explicitly represents a scene with a set of anisotropic Gaussians, whose number often exceeds millions in real-world scenarios \cite{papantonakis2024reducing}. To avoid excessive memory footprint, the rendering of each frame is typically split into tiles, so that only the Gaussians that might contribute to the current tile are accessed instead of the entire set. However, the \textbf{large tile size} and \textbf{over-inclusive Gaussian testing} cause pixels to process a substantial number of unnecessary Gaussians \cite{lee2024gscore}. This not only wastes computational resources, but more importantly, disrupts the efficient alignment of dataflow across parallel hardware units, leading to low hardware utilization. For edge devices \cite{meta2023quest3} with limited computing units, such inefficiency is particularly detrimental to both performance and energy efficiency.

\begin{figure}[htbp]
    \centering
    \includegraphics[width=1\linewidth]{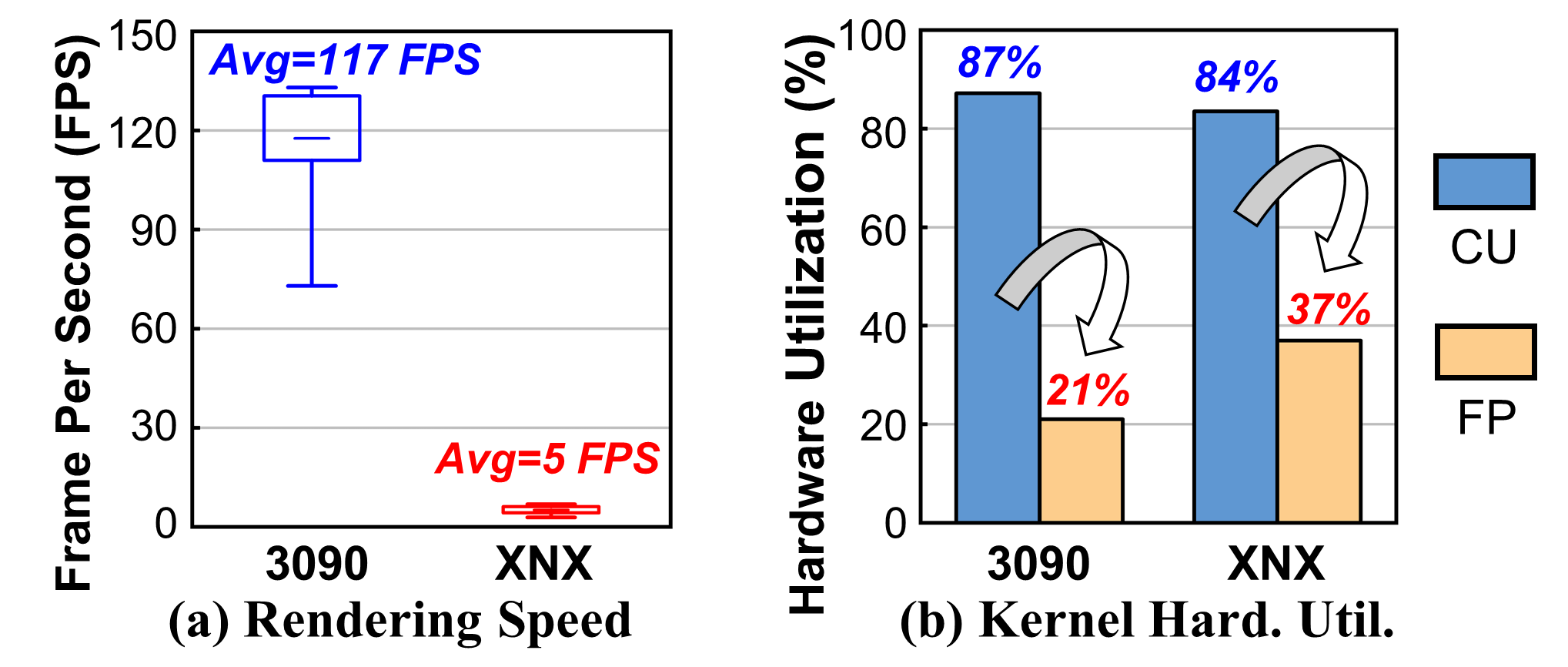}
    \caption{(a) Rendering speed and (b) hardware utilization of the rendering kernel in vanilla 3DGS, profiled on RTX 3090 \cite{nvidia2023rtx3090} and Jetson Xavier NX \cite{nvidia2023jetsonxavier}. In (b), CU denotes the utilization of compute units (i.e., GPU SMs), reflecting overall computation activity, while FP indicates the achieved FP32 performance relative to the device peak performance.}

    \label{Fig1_GPU_hardware_profiling}
    \vspace{-0.3cm}
\end{figure}

To address the aforementioned challenge, several works \cite{lee2024gscore,hanson2025speedy,wang2024adr,feng2025flashgs} try to refine the identification of contributing Gaussians by considering more Gaussian features. These features help narrow the candidate contribution region of Gaussians, thereby reducing the pixels that unnecessarily process them. Nevertheless, striving to make the candidate region exactly match the true contribution area will inevitably incurs significant computational complexity. Instead, \cite{huang2025seele} adopts a GPU-based Contribution-Aware Test (CAT) by directly evaluating each Gaussian’s contribution to a leader pixel within a pixel group before rendering. If the contribution is negligible, the entire pixel group can skip processing that Gaussian. This method alone achieves a $1.3\times$ speedup for the overall system.

Despite its effectiveness, applying this contribution-aware strategy to edge designs faces \textbf{several challenges}. \textbf{First}, the overhead of CAT is substantial. For instance, the $2\times2$ pixel group size adopted in \cite{huang2025seele} incurs significant computational overhead, as each four-pixel group requires a contribution test of a leader pixel, and the total number of such tests scales linearly with image resolution, making it prohibitively expensive for edge devices. \textbf{Second}, enabling independent Gaussian skipping for each pixel group requires dedicated memory allocation per group, which will lead to significant on-chip memory overhead. \textbf{Third}, the leader-pixel-based approach makes rendering quality highly sensitive to pixel group size. Simply reducing the number of leader pixels by adopting larger pixel groups will significantly increase the risk of missing contributing Gaussians, resulting in noticeable image degradation.

To address the above challenge, this work make the following contribution:

\begin{itemize}
\item We propose FLICKER, a fine-grained contribution-aware accelerator that leverages hardware-software co-design to achieve accurate skipping of non-contributing Gaussians at nearly pixel-scale granularity, facilitating real-time 3DGS rendering on the edge.
    
\item We introduce an adaptive leader-pixel scheme that dynamically reduces the number of leader pixels based on Gaussian shapes. Furthermore, we propose a novel batch processing technique that organizes leader pixels into rectangular groups. By sharing intermediate results within each group, the overhead is nearly halved without compromising image quality (Sec.~\ref{Sec:Algorithm Optimization}).
    
\item We introduce a two-stage hierarchical testing flow that effectively filters Gaussians with reduced computation and memory overhead. Moreover, we design a mixed-precision CAT engine, tightly integrated with the rendering pipeline, to minimize area overhead and effectively hide CAT latency (Sec.~\ref{Sec: Hardware Architecture}).
    
\item Experimental results show that FLICKER achieves up to $1.5\times$ speedup and $2.6\times$ energy efficiency compared to the baseline design, while requiring $14\%$ less area (Sec.~\ref{sec:evaluation}).

\end{itemize}

\section{Background and Motivation}
\label{Sec:Background}
\subsection{3DGS with Contribution-Aware Rendering}
\begin{figure}[htbp]
    \centering
    \includegraphics[width=1\linewidth]{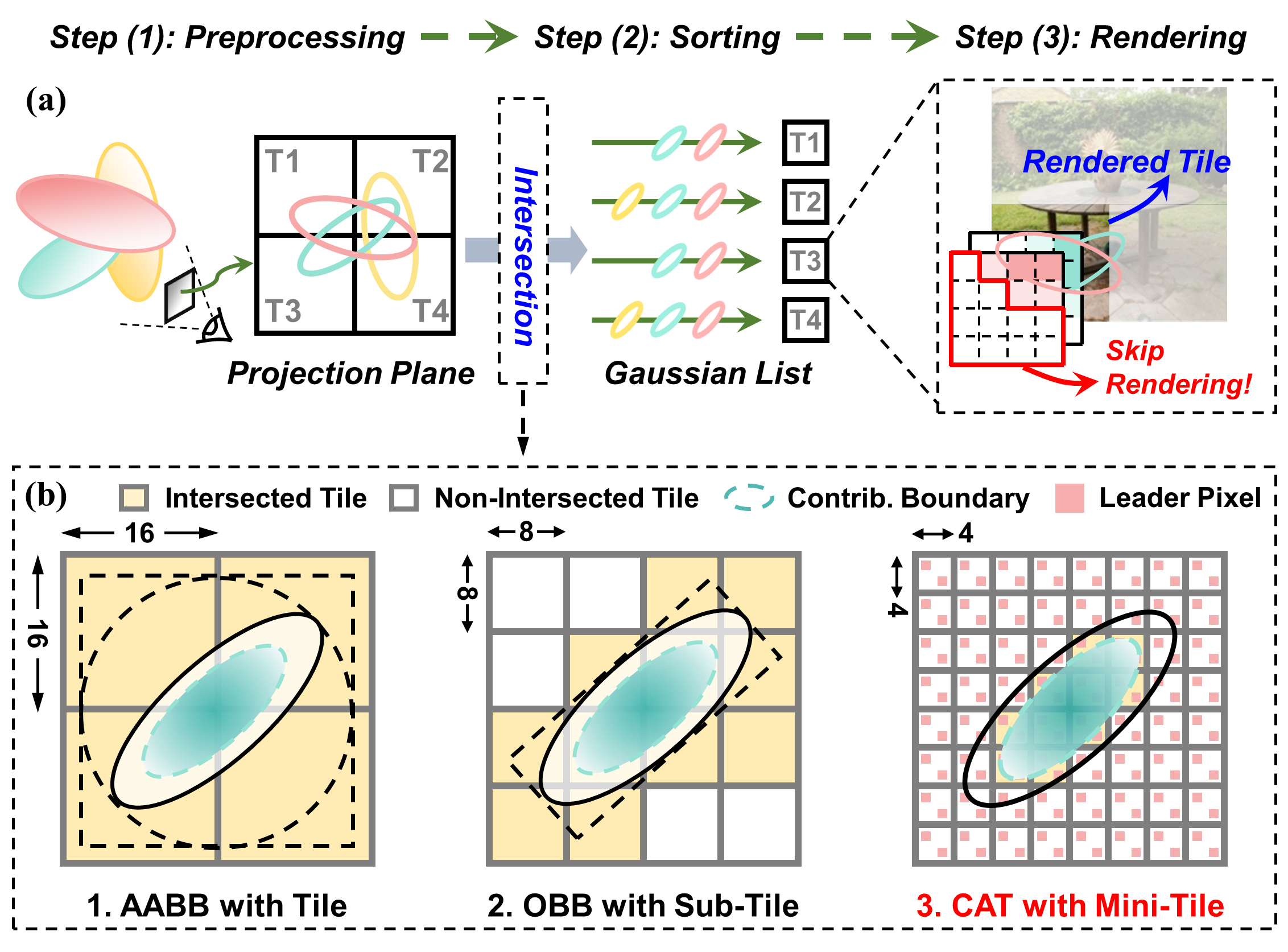}
    \caption{(a) Overall rendering pipeline of 3DGS \cite{kerbl20233d}, and (b) comparison of three intersection methods—AABB in vanilla 3DGS, OBB in GSCore \cite{lee2024gscore} and proposed Mini-Tile CAT.}
    \label{Fig2_3DGS_flow_and_intersection_method}
    \vspace{-0.3cm}
\end{figure}

\textbf{3DGS Rendering Pipeline}. The 3DGS rendering process begins with a set of Gaussian ellipsoids defined by differentiable parameters. For a given camera pose, generating an image frame from these Gaussians involves three main steps, as shown in Fig.~\ref{Fig2_3DGS_flow_and_intersection_method}(a). In \textbf{Step(1)}, Gaussians within the view frustum are projected onto the image plane, generating 2D features such as mean (\(\mu'\)), covariance (\(\Sigma'\)), and color (\(c'\)). Since rendering proceeds tile by tile, an intersection test is performed for each tile to identify Gaussians that may contribute. These Gaussians are then copied as needed to form a dedicated list for each tile. In \textbf{Step (2)}, the Gaussians in each list are then sorted by their distance from the camera (i.e., depth), arranged from near to far. With the sorted list (containing \(N\) Gaussians), all pixels within a tile are rendered in a uniform manner by iterating over the Gaussians in the list (\textbf{Step (3)}). For each Gaussian \(G_i\), the ”contribution” to pixels is first computed:

\begin{equation}
\begin{aligned}
\alpha_i = o_i \cdot e^{-\frac{1}{2}(p - \mu_i')^\top \Sigma_i^{'-1} (p - \mu_i')}
\end{aligned}
\label{Eq:alpha}
\end{equation}

where \(o_i \in [0, 1]\) is the opacity and \(p\) is the pixel coordinate. Gaussians with \(\alpha_i < \frac{1}{255}\) are considered \underline{no contribution} and skipped. Otherwise, \(\alpha_i\) is further used to compute the pixel color \( c = \sum_{i}^{N} T_i c_i' \alpha_i \). With transmittance defined as \(T_i = \prod_{j=1}^{i-1} (1 - \alpha_j)\), rendering of the current tile can terminate early if the transmittance of all pixels falls below a predefined threshold.

\textbf{Bounding Box Intersection Test}. Over-inclusive intersection tests impose significant rendering overhead. As shown in Fig.~\ref{Fig2_3DGS_flow_and_intersection_method}(b), vanilla 3DGS employs a simple Axis-Aligned Bounding Box (AABB) test \cite{kerbl20233d}. For Gaussians with anisotropic shapes, the $3\sigma$ rule defines their effective boundaries, which are then replaced by a bounding box aligned with the coordinate axes. In this toy example, the AABB test marks all tiles as intersected, leading to substantial redundant computation. GSCore \cite{lee2024gscore}, an ASIC-level 3DGS accelerator, adopts an Oriented Bounding Box (OBB) technique to better fit the anisotropic shapes of Gaussians. Moreover, by dividing tiles into sub-tiles with \(8 \times 8\) pixels, the intersected region is significantly reduced. Nevertheless, the region still does not precisely match the Gaussian's actual contribution.

\textbf{ Mini-Tile Contribution-Aware Test.} Building upon the CAT \cite{huang2025seele} discussed in Sec.~\ref{Sec:Introduction}, we propose Mini-Tile CAT, which uses multiple leader pixels to accurately capture Gaussian contributions within a $4 \times 4$ mini-tile. A mini-tile is marked intersected if any leader pixels is contributed by the Gaussian. Combined with customized optimization, this method can enable accurate intersection while largely reducing CAT overhead (details in Sec.~\ref{Sec:Algorithm Optimization}). As shown in Fig.~\ref{Fig2_3DGS_flow_and_intersection_method}(b), the intersected region closely aligns with the Gaussian's true contribution boundary, thereby ensuring higher rendering efficiency.

\subsection{3DGS Profiling and Strategy Analysis}

\begin{figure*}[t]
    \centering
    \includegraphics[width=1\textwidth]{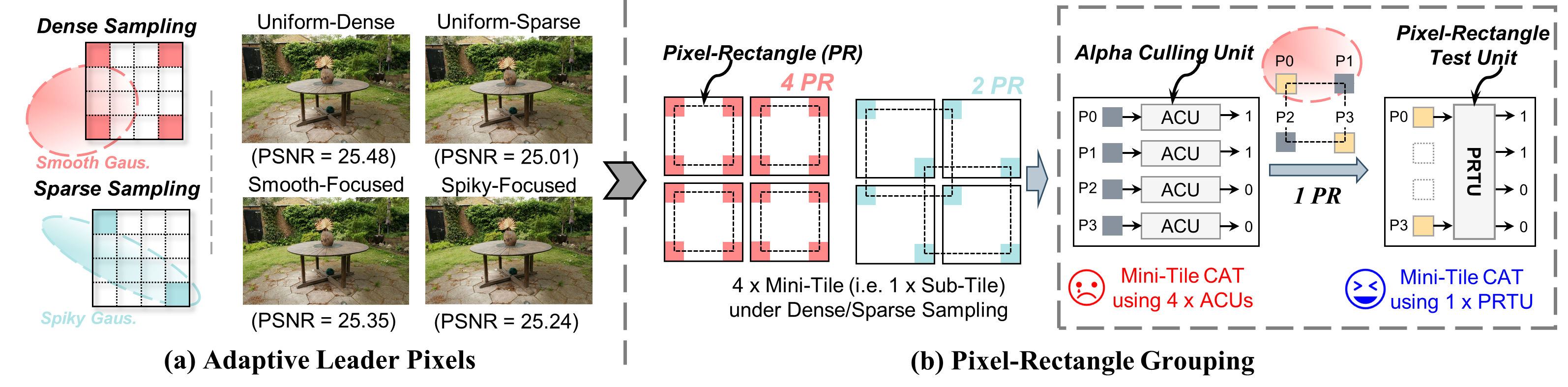}
    \caption{Mini-Tile CAT algorithm optimization: (a) adaptive leader pixels, and (b) pixel-rectangle grouping. In (a), the PSNR of vanilla 3DGS is 25.56, while the Uniform-Dense mode shows negligible loss. Although smooth Gaussians account for 43\%, the Smooth-Focused mode achieves higher PSNR, indicating that the contribution of smooth Gaussians is more significant in this case. In (b), the Mini-Tile CAT of a Pixel Rectangle (PR) can be simplified by exploiting its coordinate symmetry.}

    \vspace{-0.3cm}

\label{Fig3_Algorithm_Optimization}
\end{figure*}

We adopt a \underline{two-stage analysis method} to investigate performance bottlenecks and identify optimization opportunities: \textbf{First}, we profile the 3DGS rendering pipeline on two GPUs: a desktop-level GPU, RTX 3090~\cite{nvidia2023rtx3090}, and an edge GPU, XNX~\cite{nvidia2023jetsonxavier}. Since XNX shares similar hardware specifications to edge GPUs used in advanced VR headsets~\cite{zhao2023instant}, its profiling results can better reflect the bottlenecks in practical edge scenarios. We obtain a detailed GPU performance breakdown using Nsight Compute~\cite{nvidia_nsight_compute} and profiling is conducted with datasets from Mip-NeRF360~\cite{barron2022mip}. \textbf{Second}, we further conduct an in-depth analysis of the introduced intersection methods to quantify the expected benefits and overheads of the Mini-Tile CAT, thereby guiding the subsequent design.

\textbf{3DGS Profiling Result}. As shown in Fig.~\ref{Fig1_GPU_hardware_profiling}(a), 3DGS achieves real-time rendering on desktop GPUs, with average FPS exceeding 100. In contrast, on edge GPUs, the FPS drops sharply to around 5 per scene, highlighting such edge devices struggle to handle the 3DGS’s workload. To understand this discrepancy, we further analyzed the \underline{hardware utilization} of the rendering step in 3DGS, which accounts for a significant portion of GPU kernel execution time, often exceeding 60\% \cite{lee2024gscore,jo2025ps,wang2025famers}. As shown in Fig.~\ref{Fig1_GPU_hardware_profiling}(b), the overall compute unit utilization on the GPU, calculated via SM Core throughput~\cite{jia2018dissecting}, reaches an average of 85\%. However, the average floating-point utilization is merely 29\%, which corresponds to the core computations in the rendering step. The result indicates that the already limited computation resources on edge GPUs are underutilized, exacerbating performance degradation. This inefficiency stems from the fact that, during rendering, certain pixels within the same tile are skipped due to their negligible Gaussian contributions, which causes warp divergence. Moreover, the over-inclusive intersection test further amplifies this effect, ultimately leading to poor performance.

\begin{wrapfigure}{r}{0.45\linewidth} % r表示右侧放置, 0.4\linewidth 控制宽度
    \vspace{-0.4cm}
    \centering
    \includegraphics[width=1\linewidth]{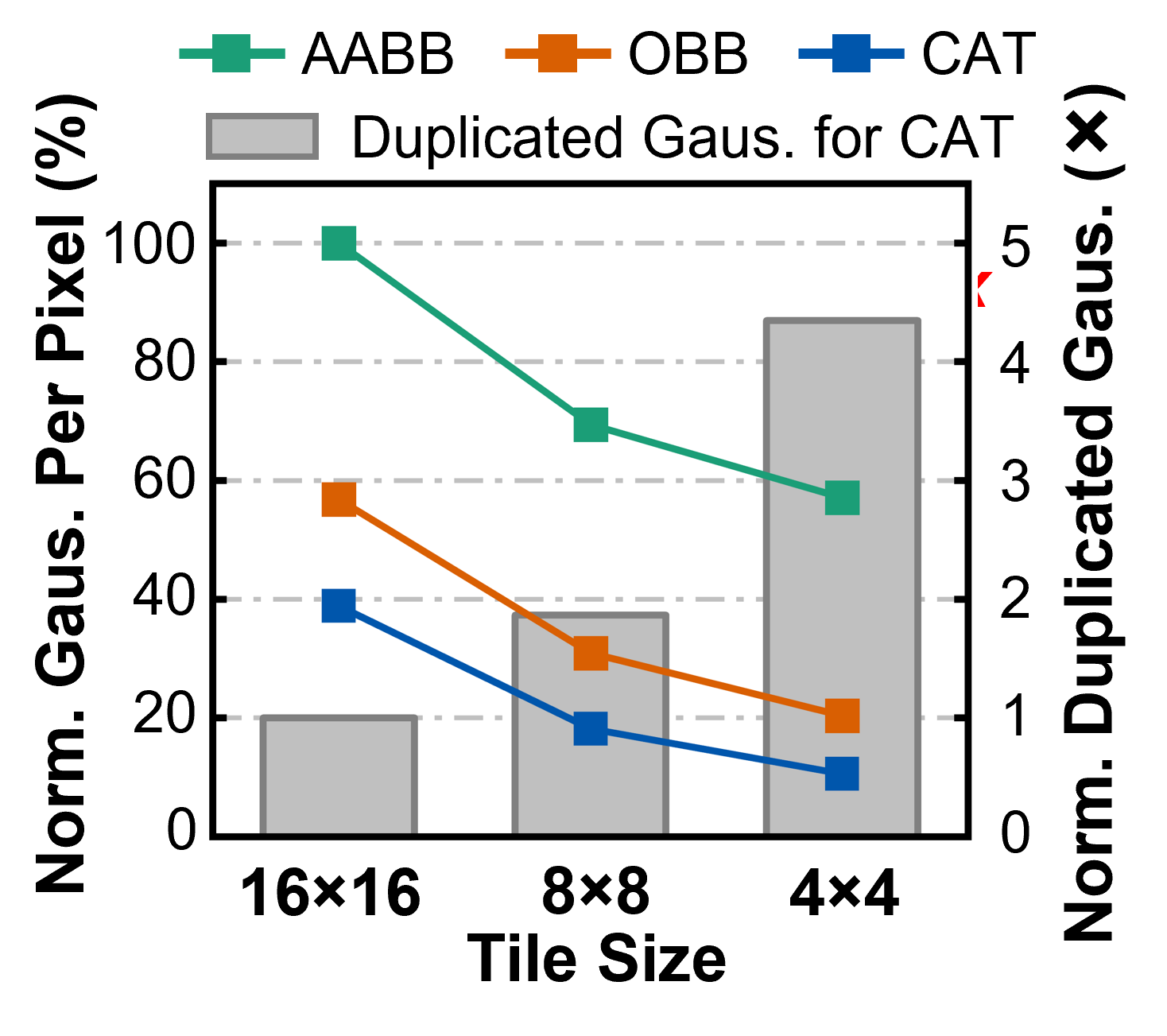} 
    \caption{Per-pixel processed Gaussians across intersection strategies and duplicate Gaussians across tile sizes.}

    \label{Fig4_Method_profiling}
\end{wrapfigure}

\textbf{Strategy Analysis}.
Based on profiling, we further compare introduced intersection methods on a real-world scene to guide Mini-Tile CAT implementation on edge hardware. In this analysis, each mini-tile is initially assigned 4 leader pixels. The results, shown in Fig.~\ref{Fig4_Method_profiling}, lead to the \underline{following observations}: \textbf{First}, Mini-Tile CAT demonstrates a clear advantage in filtering Gaussians at the fine-grained $4 \times 4$ tile level, i.e., mini-tile. Compared to AABB on a $16 \times 16$ tile, Mini-Tile CAT reduces the number of Gaussians each pixel must process to only 10\%, the lowest among all tested methods. This underscores its potential to substantially cut rendering workload and alleviate underutilization of parallel hardware units caused by pixel skipping. \textbf{Second}, although Mini-Tile CAT reduces unnecessary Gaussian processing, it incurs significant computational overhead because leader pixels must be tested on many Gaussians. Moreover, since Mini-Tile CAT evaluates far more Gaussians than are ultimately forwarded to the rendering stage, it can easily become a pipeline bottleneck, causing downstream stalls. This highlights the need for an efficient architectural support of Mini-Tile CAT. \textbf{Third}, smaller tile sizes more effectively reduce redundant Gaussian processing but at the cost of higher memory overhead due to increased duplicated Gaussians. For instance, reducing the tile size from $16 \times 16$ to $4 \times 4$ pixels increases the total number of duplicated Gaussians to $4\times$ the original, highlighting the need for an efficient hierarchical intersection strategy and memory allocation.

\section{Algorithm Optimization}
\label{Sec:Algorithm Optimization}
In this section, we introduce two algorithmic strategies to effectively reduce the computation overhead of Mini-Tile CAT from two perspectives: reducing (1) the number of required leader pixels, and (2) the per-leader-pixel CAT overhead.

\vspace{-2mm}
\subsection{Adaptive Leader Pixels}
\textbf{Basic Mode}. We begin by defining two sampling settings for Mini-Tile CAT: \textit{Dense Sampling}, using four corner pixels per mini-tile, and \textit{Sparse Sampling}, using two diagonal pixels. From these, we can derive two modes for Mini-Tile CAT: \textit{Uniform-Dense}, where all Gaussians use Dense Sampling, and \textit{Uniform-Sparse}, where all use Sparse Sampling. As shown in Fig.~\ref{Fig3_Algorithm_Optimization}(a), Uniform-Dense captures most contributing Gaussians with only a negligible PSNR drop (0.08 dB) compared to vanilla 3DGS, while Uniform-Sparse, though reducing leader pixels by half, causes noticeable quality degradation.

\textbf{Adaptive Mode}. To combine the strengths of both uniform modes, we propose an adaptive strategy that dynamically switches the sampling based on Gaussian shape. All Gaussians are first classified as \textit{Smooth} ($\text{axis ratio} < 3$) or \textit{Spiky} ($\text{axis ratio} \geq 3$). Smooth Gaussians use Dense Sampling for broader coverage (\textit{Smooth-Focused} mode), while spiky Gaussians use Sparse Sampling to save leader pixels. As shown in Fig.~\ref{Fig3_Algorithm_Optimization}(a), this adaptive mode reduces PSNR loss by 73\% compared to Uniform-Sparse, while retaining 57\% of its leader-pixel savings. Notably, if spiky Gaussians carry more critical visual details, the strategy can be switched to a \textit{Spiky-Focused} mode, in which Dense Sampling is applied to spiky Gaussians.

\vspace{-2mm}
\subsection{Pixel-Rectangle Grouping}

\textbf{CAT for Single Pixels.} As discussed in Sec.~\ref{Sec:Background}, the contribution of a Gaussian to a leader pixel is calculated using Eq.~\ref{Eq:alpha}. The result is then compared to a threshold (e.g., $\alpha < \frac{1}{255}$) to determine whether the Gaussian can be skipped. A straightforward way for CAT is to design a dedicated Alpha Culling Unit (ACU) \cite{lee2024gscore,wang2025famers,jo2025ps} to test each pixel individually. However, this approach incurs significant computation overhead, especially when testing multiple leader pixels across dense mini-tiles.

\textbf{CAT for Pixel Rectangles.} To reduce the average CAT overhead for per leader pixel, we first simplify the Eq.~\ref{Eq:alpha} as follows:
\begin{equation}
\ln(255 \cdot o) > -\frac{1}{2}(p - \mu')^\top \Sigma^{'-1} (p - \mu')
\label{eq:alpha_simplified_test}
\end{equation}

In the inequality, the left-hand side term, $\ln(255 \cdot o)$, is identical for all leader pixels tested against the same Gaussian, and therefore only needs to be computed once and shared. To reduce the overhead for computing the right-hand side, i.e., the Gaussian Weight $E$, we propose a \textit{Pixel-Rectangle Test Unit (PRTU)}, which evaluates the contribution of a Gaussian to a group of four leader pixels arranged in a rectangular \textit{Pixel Rectangle (PR)}. Within each PR, the two off-diagonal corner pixels have symmetric coordinates relative to the main-diagonal pixels, allowing the intermediate results from the main-diagonal pixels to be reused for computing the off-diagonal pixels. The pseudocode for processing a PR is provided in Alg.~\ref{Alg:pixel rectangle}. 

\begin{algorithm}
\renewcommand{\algorithmicrequire}{\textbf{Input:}}
\renewcommand{\algorithmicensure}{\textbf{Output:}}
\caption{Pixel-Rectangle Gaussian Weight Computation}
\label{algorithm1}
\begin{algorithmic}[1]
\REQUIRE Gaussian mean $\mu'$, conic entries $\Sigma_{xx}'^{-1}, \Sigma_{yy}'^{-1}, \Sigma_{xy}'^{-1}$, Main diagonal pixel coordinates $\mathbf{p}_\text{top}, \mathbf{p}_\text{bot}$ (correspond to $p_0$ and $p_3$ in a PR).

\ENSURE Gaussian weight $E_0, E_1, E_2, E_3$
\STATE $\Delta_\text{top} \gets \mathbf{p}_\text{top} - \mu'$ \quad $\Delta_\text{bot} \gets \mathbf{p}_\text{bot} - \mu'$

\STATE $s^x_\text{top} = 0.5 \cdot \Delta_{\text{top},x}^2 \cdot \Sigma_{xx}'^{-1}$ \quad
       $s^y_\text{top} = 0.5 \cdot \Delta_{\text{top},y}^2 \cdot \Sigma_{yy}'^{-1}$
\STATE $s^x_\text{bot} = 0.5 \cdot \Delta_{\text{bot},x}^2 \cdot \Sigma_{xx}'^{-1}$ \quad
       $s^y_\text{bot} = 0.5 \cdot \Delta_{\text{bot},y}^2 \cdot \Sigma_{yy}'^{-1}$

\STATE $t_0 = \Delta_{\text{top},x} \cdot \Delta_{\text{top},y} \cdot \Sigma_{xy}'^{-1}$ \quad
       $t_1 = \Delta_{\text{bot},x} \cdot \Delta_{\text{top},y} \cdot \Sigma_{xy}'^{-1}$
\STATE $t_2 = \Delta_{\text{top},x} \cdot \Delta_{\text{bot},y} \cdot \Sigma_{xy}'^{-1}$ \quad
       $t_3 = \Delta_{\text{bot},x} \cdot \Delta_{\text{bot},y} \cdot \Sigma_{xy}'^{-1}$

\STATE $E_0 = s^x_\text{top} + s^y_\text{top} + t_0$ \quad
       $E_1 = s^x_\text{bot} + s^y_\text{top} + t_1$
\STATE $E_2 = s^x_\text{top} + s^y_\text{bot} + t_2$ \quad
       $E_3 = s^x_\text{bot} + s^y_\text{bot} + t_3$
\end{algorithmic}
\label{Alg:pixel rectangle}
\end{algorithm}
Compared to the ACU that tests pixels individually, our pixel-rectangle grouping method nearly halves the computation cost. Most importantly, it can be effectively combined with our adaptive leader-pixel strategy. As shown in Fig.~\ref{Fig3_Algorithm_Optimization}(b), a sub-tile composed of four mini-tiles typically includes multiple PRs: in Dense Sampling, each mini-tile contributes one PR, resulting in four PRs per sub-tile, whereas Sparse Sampling still can form two valid PRs across mini-tiles.

\section{Hardware Architecture}
\label{Sec: Hardware Architecture}
We begin with an overview of the FLICKER architecture, then detail the hierarchical Gaussian testing and contribution-aware rendering pipeline, which enables high-throughput Mini-Tile CAT through dedicated architectural support. Finally, we introduce the mixed-precision contribution-aware test unit (CTU) for applying Mini-Tile CAT with minimal hardware overhead.

\subsection{Overall Architecture} 
\begin{figure}[htbp]
    \centering
    \includegraphics[width=1\linewidth]{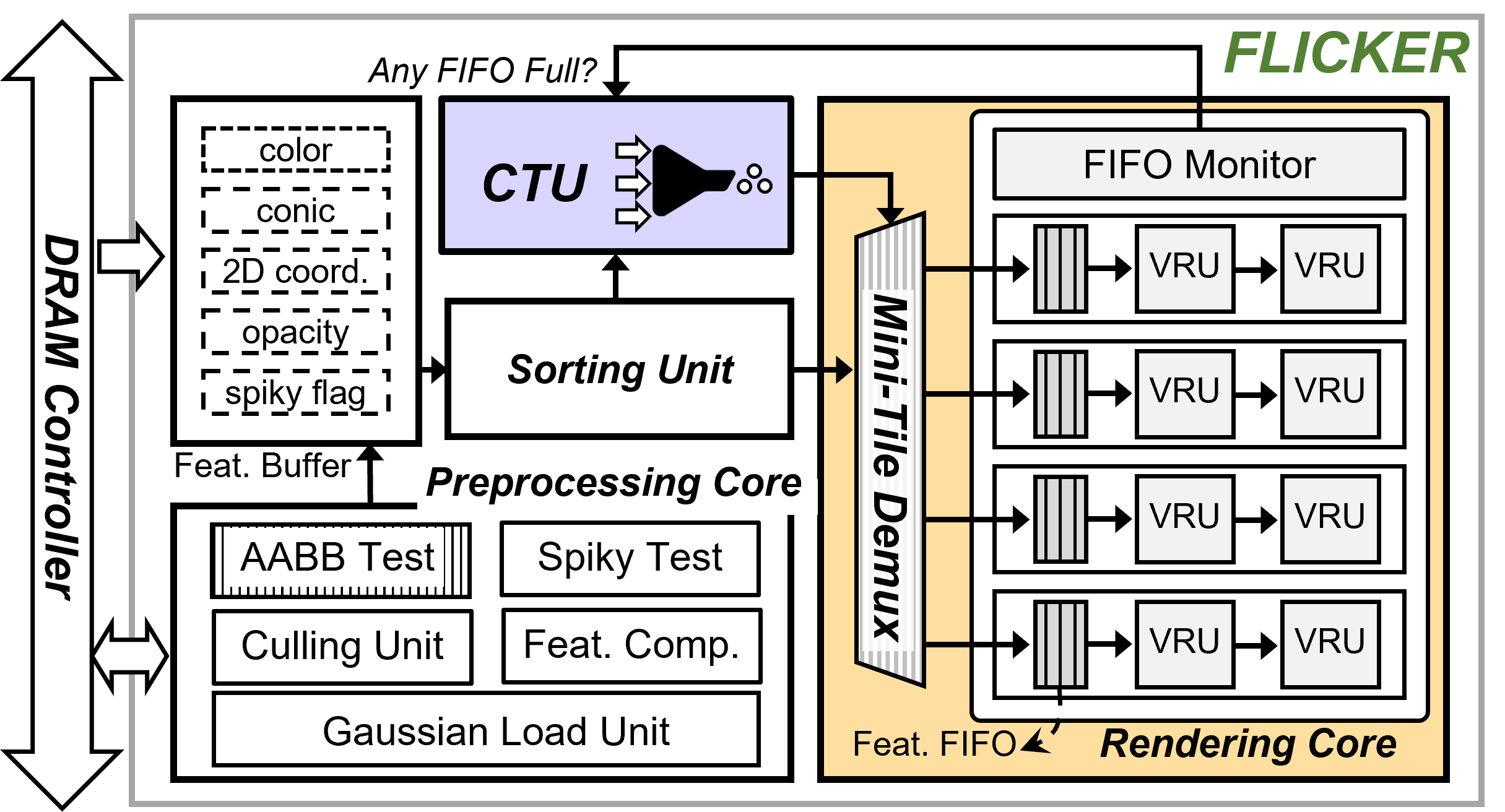}
    \caption{Overall hardware architecture of FLICKER. The key component, the contribution-aware test unit (CTU), is highlighted in purple and will be detailed in Sec.~\ref{sec:mixed precision contribution aware test unit}.}
\label{Fig5_Overall_architecture}
\vspace{-0.3cm}
\end{figure}

\textbf{Main Components}. As shown in Fig.~\ref{Fig5_Overall_architecture}, the architecture consists of four main components: preprocessing core, sorting unit, CTU, and rendering core. The preprocessing core projects 3D Gaussian features into 2D, determines whether Gaussians fall within the frustum, classifies them as spiky or smooth, and performs AABB tests for sub-tile intersections. The sorting unit fetchs the converted features, sorts them by depth, and forwards them to the CTU. The CTU applies Mini-Tile CAT to filter sorted Gaussians according to their contribution. Finally, the rendering core completes the rendering step using the Gaussians that pass CTU.

\textbf{Memory Access Optimization}. Since the number of Gaussians is extremely large, most parameters must be stored off-chip. To reduce DDR traffic, we adopt a clustering method that groups multiple Gaussians into larger “big Gaussians” \cite{jo2025ps}. Frustum culling is then performed on these big Gaussians instead of on individual ones, significantly reducing the number of DDR accesses for the preprocessing core. Moreover, the bandwidth efficiency is further improved by loading only geometric features (10 parameters) during culling, while color features (45 parameters) and other parameters are fetched only for Gaussians that pass frustum culling and intersection test.

\subsection{Hierarchical Gaussian Testing and Contribution-Aware Rendering Pipeline}
\begin{figure}[htbp]
    \centering
    \includegraphics[width=0.9\linewidth]{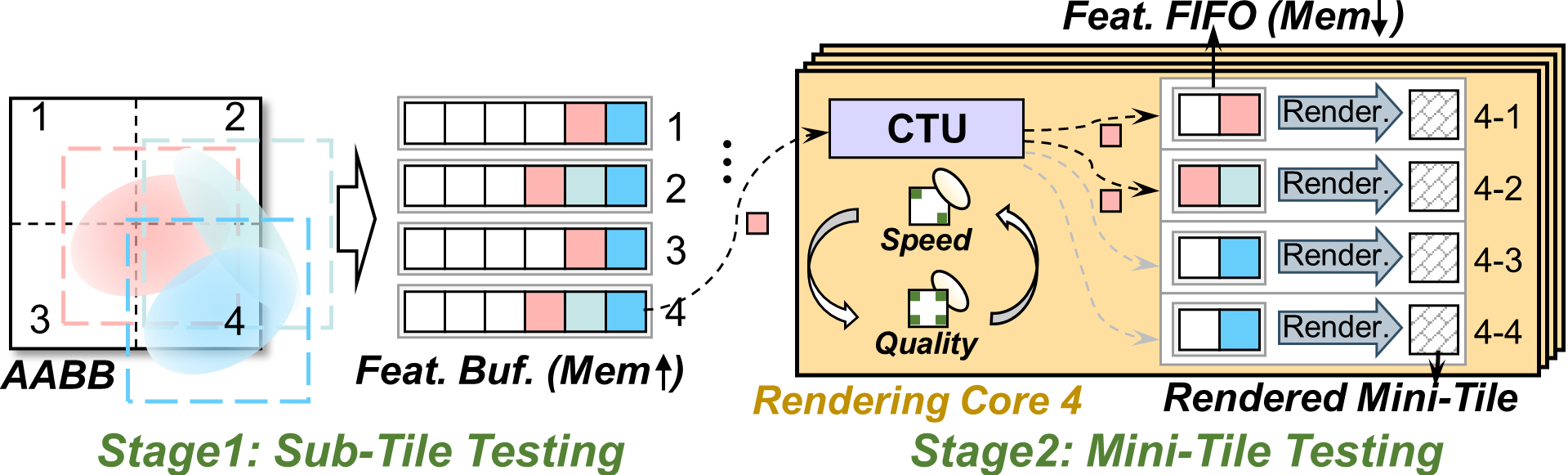}
    \caption{Hierarchical Gaussian testing.}
    \label{Fig6_Hierachy_Testing_and_Rendering_Pipeline}
    \vspace{-0.2cm}
\end{figure}

\textbf{Hierarchical Testing}. As discussed in Sec.~\ref{Sec:Algorithm Optimization}, Mini-Tile CAT reduces per-pixel overhead, but the number of Gaussians to test remains high. To handle this, we introduce a two-stage hierarchical testing strategy (Fig.~\ref{Fig6_Hierachy_Testing_and_Rendering_Pipeline}). \textbf{Stage 1:} In the preprocessing core, a sub-tile AABB test is performed. Gaussians are duplicated into feature buffers according to their sub-tile intersection mask, enabling efficient skipping at the sub-tile level while reducing the CTU workload (by 30\%, as shown in Fig.~\ref{Fig4_Method_profiling}). \textbf{Stage 2:} The CTU processes Gaussians that pass Stage 1 by applying Mini-Tile CAT to generate fine-grained masks. Based on these masks, the contributing Gaussians are then duplicated into the corresponding FIFOs in the rendering core. Each FIFO drives two VRUs, which together render 16 pixels—exactly one mini-tile. Four such channels within a rendering core cover one sub-tile, while the four rendering cores in FLICKER collectively span a full tile. Since hierarchical testing reduces the Gaussian count to about 10\% of the original, the required FIFO capacity is small, which in turn lowers memory overhead. Overall, this organization enables efficient and fine-grained mini-tile skipping under the tile level.

\textbf{Contribution-Aware Rendering Pipeline}. Beyond hierarchical testing, we further optimize the runtime pipeline to ensure smooth execution. With the dedicated CTU for Mini-Tile CAT, most of its latency is hidden by overlapping with VRU rendering. To further reduce FIFO capacity, we design a stall-resilient pipeline. When any FIFO inside the rendering core becomes full, a FIFO monitor detects the stall and notifies the CTU (Fig.~\ref{Fig5_Overall_architecture}). Upon receiving the stall signal, the CTU halts the intake of new Gaussians, while the in-flight pipeline results are safely stored in a small built-in CTU FIFO, ensuring no data loss despite its fully pipelined nature. As validated in Sec.~\ref{subsec:critical_component_analysis}, this design allows very shallow FIFOs to achieve most of the speedup provided by mini-tile skipping. Instead, if CTU throughput falls behind the VRUs, the system can switch to Uniform-Sparse mode, boosting Mini-Tile CAT throughput.

\subsection{Mixed-Precision Contribution-Aware Test Unit}
\label{sec:mixed precision contribution aware test unit}
\begin{figure}[htbp]
    \vspace{-0.3cm}
    \centering
    \includegraphics[width=1\linewidth]{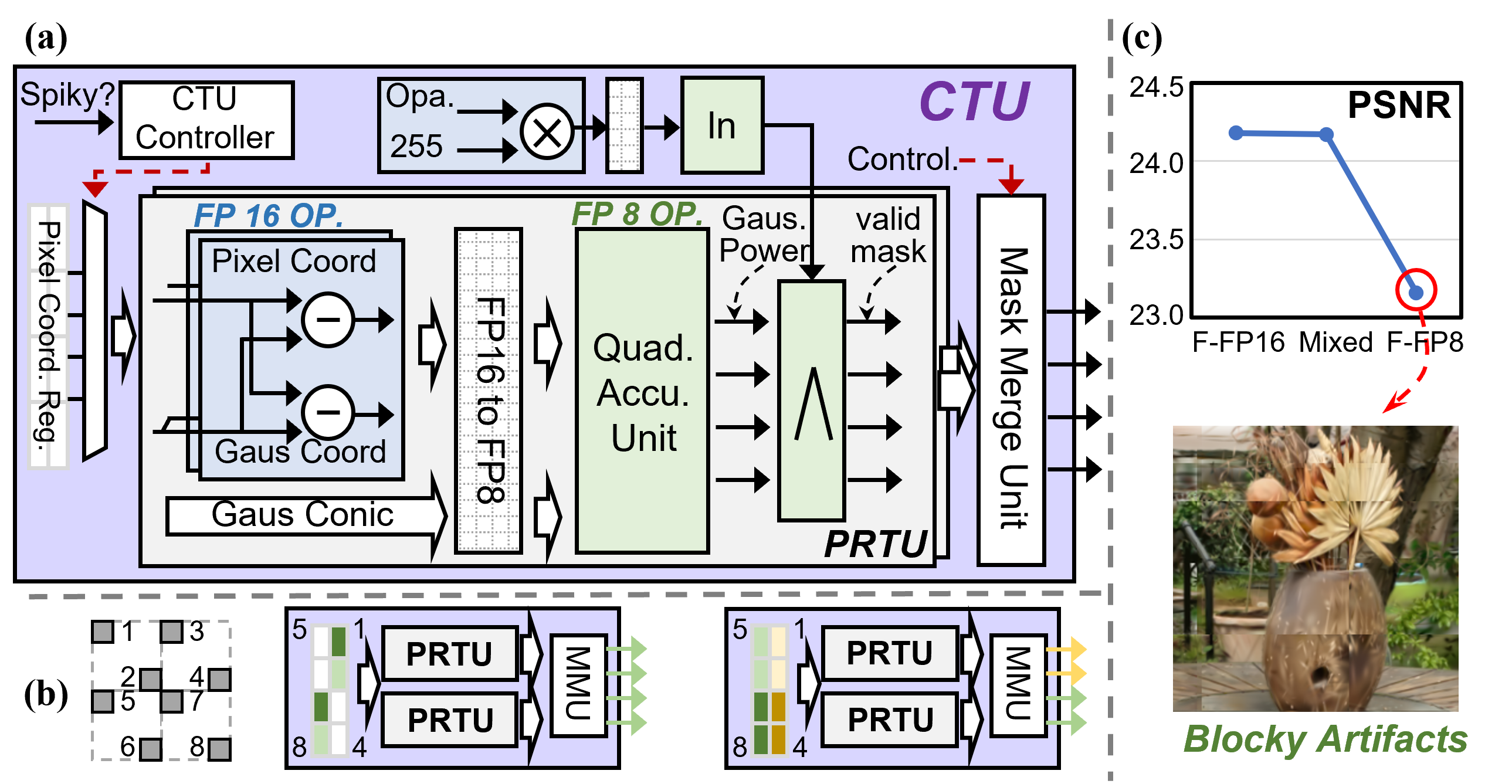}
    \caption{Mixed-precision contribution test unit: (a) microarchitecture, (b) dataflow for the adaptive leader pixel test, and (c) comparison across different precision schemes.}

    \label{Fig7_Mixed_Precision_Contribution_Test_Unit}

    \vspace{-0.3cm}
\end{figure}

As shown in Fig.~\ref{Fig7_Mixed_Precision_Contribution_Test_Unit}(a), the mixed-precision CTU architecture consists of two PRTUs, a Mask Merge Unit (MMU), and units for computing the shared term \(\ln(255 \cdot o)\). The CTU is fully pipelined and can process two PRs (total 8 leader pixels) per cycle, with each PRTU handling one PR. The controller dynamically adjusts the sampling mode based on the Gaussian spiky flag. For Sparse Sampling, the two PRTUs directly generate test masks for two PRs, which are then merged by the MMU and output. For Dense Sampling, four PRs are processed in two batches: the mask from the first batch is stored in registers, and after the second batch completes, the MMU merges both batches to produce the final output, as illustrated in Fig.~\ref{Fig7_Mixed_Precision_Contribution_Test_Unit}(b).

To reduce the hardware overhead of the CTU, we employ a mixed-precision PRTU: differences between pixel and Gaussian coordinates (line 1 in Alg.~\ref{Alg:pixel rectangle}) are computed in FP16, and the results are then converted to FP8 for subsequent calculations in the Quarda Accumulation Unit (lines 2--7). We evaluate three precision schemes: Full FP16, Full FP8, and mixed precision. As shown in Fig.~\ref{Fig7_Mixed_Precision_Contribution_Test_Unit}(c), the mixed precision scheme maintains high image quality, whereas Full FP8 suffers from severe PSNR degradation and noticeable blocky artifacts. The degradation primarily arises from the compression of relative positional information between pixels and Gaussians, which leads to the loss of interpolation details. In contrast, our mixed precision scheme preserves critical interpolation information and leverages the inherent error tolerance of Mini-Tile CAT, finally ensuring quality with low hardware cost.

\section{Evaluation}
\label{sec:evaluation}
\subsection{Experimental Settings}
\textbf{Algorithm Setup}. We evaluate on eight real-world scenes: two outdoor scenes from Tanks \& Temples, four outdoor scenes from Mip-NeRF 360, and two indoor scenes from Deep Blending. Each scene is first trained with vanilla 3DGS \cite{kerbl20233d} for 30K iterations to obtain baseline models. To produce more compact models, we apply a pruning technique \cite{ali2024trimming}, which removes Gaussians with negligible contribution, followed by an additional 3K fine-tuning iterations. After pruning, we adopt the clustering method \cite{jo2025ps} to group Gaussians into clusters. Training is performed in FP32, and parameters are then quantized for full FP16 rendering on FLICKER.

\textbf{Hardware Setup}. The proposed accelerator comprises $4\times(4\times2)$ VRUs (4 rendering cores), 4 CTUs, 4 sorting units, and 4 preprocessing cores. The design is developed in Verilog and synthesized with Synopsys Design Compiler using the TSMC 28nm process, with SRAMs generated via the memory compiler. To evaluate performance, we build a cycle-accurate simulator of FLICKER, including an LPDDR4 memory with 51.2 GB/s bandwidth, and estimate DRAM energy following \cite{dong202528nm}\cite{song20151}. For comparison, we use GSCore \cite{lee2024gscore} and the Jetson XNX GPU \cite{nvidia2023jetsonxavier} as baselines. In addition, we build a simplified version of FLICKER without the CTU to assess its impact.

\begin{table}[]
\caption{Evaluation of rendering quality (PSNR$\uparrow$ and SSIM$\uparrow$) across different approaches}
\label{tab:psnr_ssim_comparison}
\begin{tabular}{c!{\vrule width 1.2pt}cc!{\vrule width 1.2pt}cc!{\vrule width 1.2pt}cc!{\vrule width 1.2pt}c}
\Xhline{1.2pt} % 加粗最上方横线
\multirow{2}{*}{}
& \multicolumn{2}{c!{\vrule width 1.2pt}}{\begin{tabular}[c]{@{}c@{}}Tanks \& \\ Temples \cite{knapitsch2017tanks}\end{tabular}} 
& \multicolumn{2}{c!{\vrule width 1.2pt}}{\begin{tabular}[c]{@{}c@{}}MipNeRF360 \\ (outdoor) \cite{barron2022mip}\end{tabular}} 
& \multicolumn{2}{c!{\vrule width 1.2pt}}{\begin{tabular}[c]{@{}c@{}}Deep \\ Blending \cite{hedman2018deep}\end{tabular}}  
& Average \\ \cline{2-8} 
& \multicolumn{1}{c|}{PSNR} & SSIM & \multicolumn{1}{c|}{PSNR} & SSIM & \multicolumn{1}{c|}{PSNR} & SSIM & PSNR \\ \hline
Base.             
& \multicolumn{1}{c|}{24.08} & 0.86 & \multicolumn{1}{c|}{25.88} & 0.76 & \multicolumn{1}{c|}{29.72} & 0.90 & 26.56 \\ \hline
Prun.             
& \multicolumn{1}{c|}{23.61} & 0.84 & \multicolumn{1}{c|}{24.71} & 0.73 & \multicolumn{1}{c|}{29.64} & 0.90 & 25.99 \\ \hline
Ours              
& \multicolumn{1}{c|}{23.51} & 0.84 & \multicolumn{1}{c|}{24.52} & 0.73 & \multicolumn{1}{c|}{29.62} & 0.90 & 25.88 \\ 
\Xhline{1.2pt} % 加粗最下方横线
\end{tabular}
\vspace{-0.4cm}
\end{table} 

\begin{figure}[htbp]
    \centering
    \includegraphics[width=0.9\linewidth]{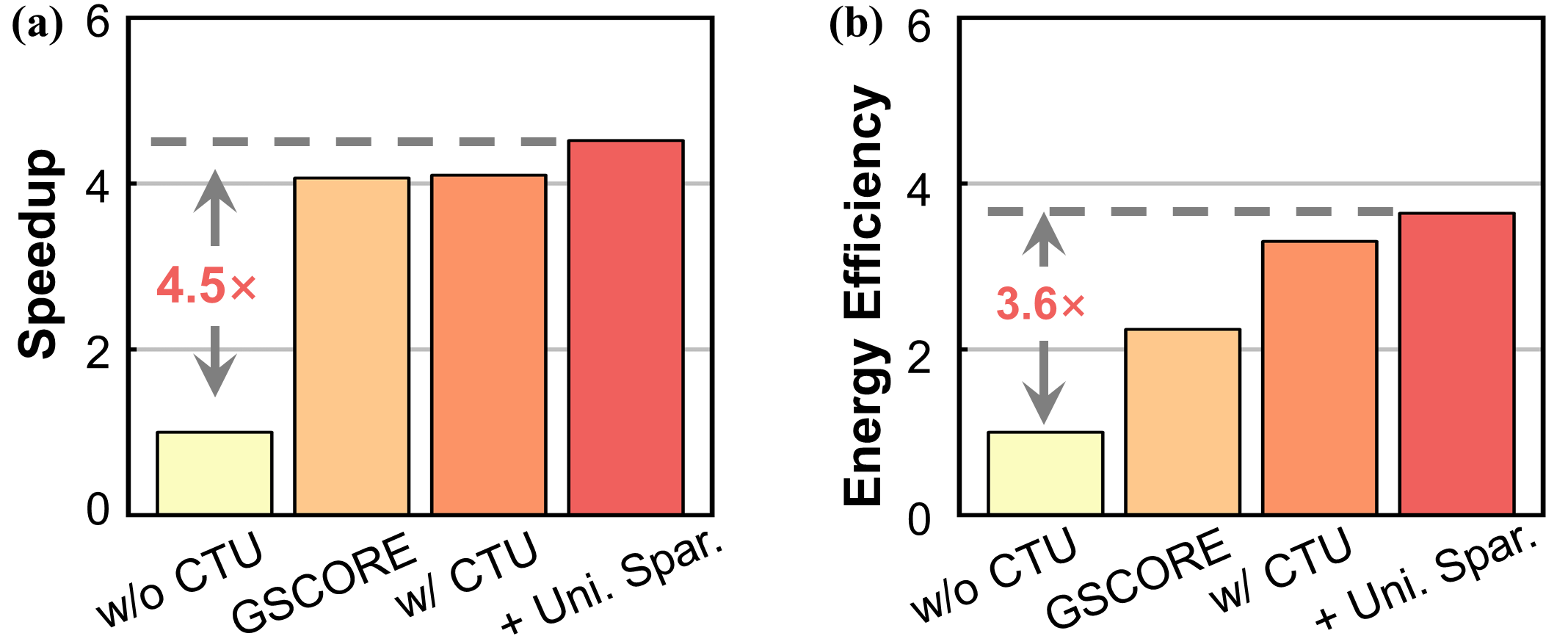}
    \caption{Comparison of (a) speedup and (b) energy efficiency for the rendering stage. Note that GSCore \cite{lee2024gscore} is configured with 64 VRUs, while ours uses 32 VRUs for smaller area. The evaluation is performed on the scene \textit{Garden} only. }
    \label{Fig8_Rendering_Stage_Evaluation}
\end{figure}

\begin{figure}[htbp]
    \vspace{-0.3cm}
    \centering
    \includegraphics[width=0.95\linewidth]{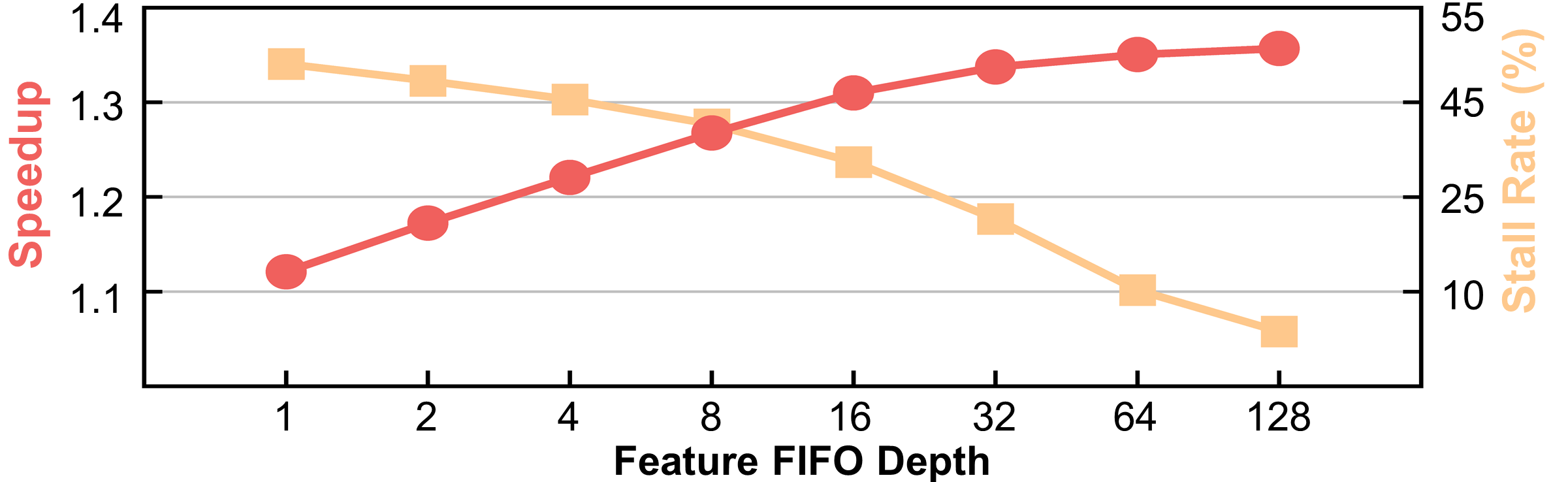}
    \caption{The sensitivity of speedup and CTU stall rate to the depth of the feature FIFO in the rendering stage. The evaluation is performed on the scene \textit{Garden} only. }
\label{Fig9_Analysis_across_FIFO_depth}
\vspace{-0.5cm}

\end{figure}

\subsection{Critical Component Analysis}
\label{subsec:critical_component_analysis}
Fig.~\ref{Fig8_Rendering_Stage_Evaluation} presents the normalized speedup and energy savings obtained from employing the CTU. To highlight its contribution, we evaluate on the baseline model without other optimizations and focus solely on the rendering stage. In terms of \underline{speedup}, the simplified version of FLICKER is 4$\times$ slower than GSCore, primarily because it only adopts a basic AABB test, whereas GSCore employs an OBB test \cite{lee2024gscore} and doubles the number of VRUs (32 vs. 64). Integrating the CTU improves performance to $4\times$ over the baseline by enabling accurate mini-tile Gaussian skipping. Even with fewer VRUs, FLICKER still matches the rendering speedup of GSCore. Further configuring the CTU in Uniform-Sparse mode yields an additional $1.1\times$ speedup, as it helps efficiently skip large groups of non-contributing Gaussians that would otherwise stall the VRUs. For \underline{energy efficiency}, our design achieves up to 1.6$\times$ energy savings over GSCore, as it prevents the massive VRUs from wasting energy on non-contributing Gaussians.

To quantify the impact of FIFO depth on CTU stalls (when FIFO is full), we evaluate the rendering-stage speedup of FLICKER across depths from 1 to 128 (Fig.~\ref{Fig9_Analysis_across_FIFO_depth}) and the corresponding CTU stall rates. Results show that increasing FIFO depth reduces stalls and improves speedup, reaching a maximum of 1.36$\times$ at depth 128. However, returns diminish beyond a depth of 16, which already achieves 96\% of the maximum speedup while using only 12.5\% of the memory compared to depth 128. Therefore, we select a FIFO depth of 16 for configuration. This highlights the effectiveness of hierarchical testing, which enables mini-tile skipping with shallow FIFOs rather than large sub-tile buffers and achieves most of the performance gain with minimal memory overhead.

\begin{figure}[htbp]
    \centering
    \includegraphics[width=0.9\linewidth]{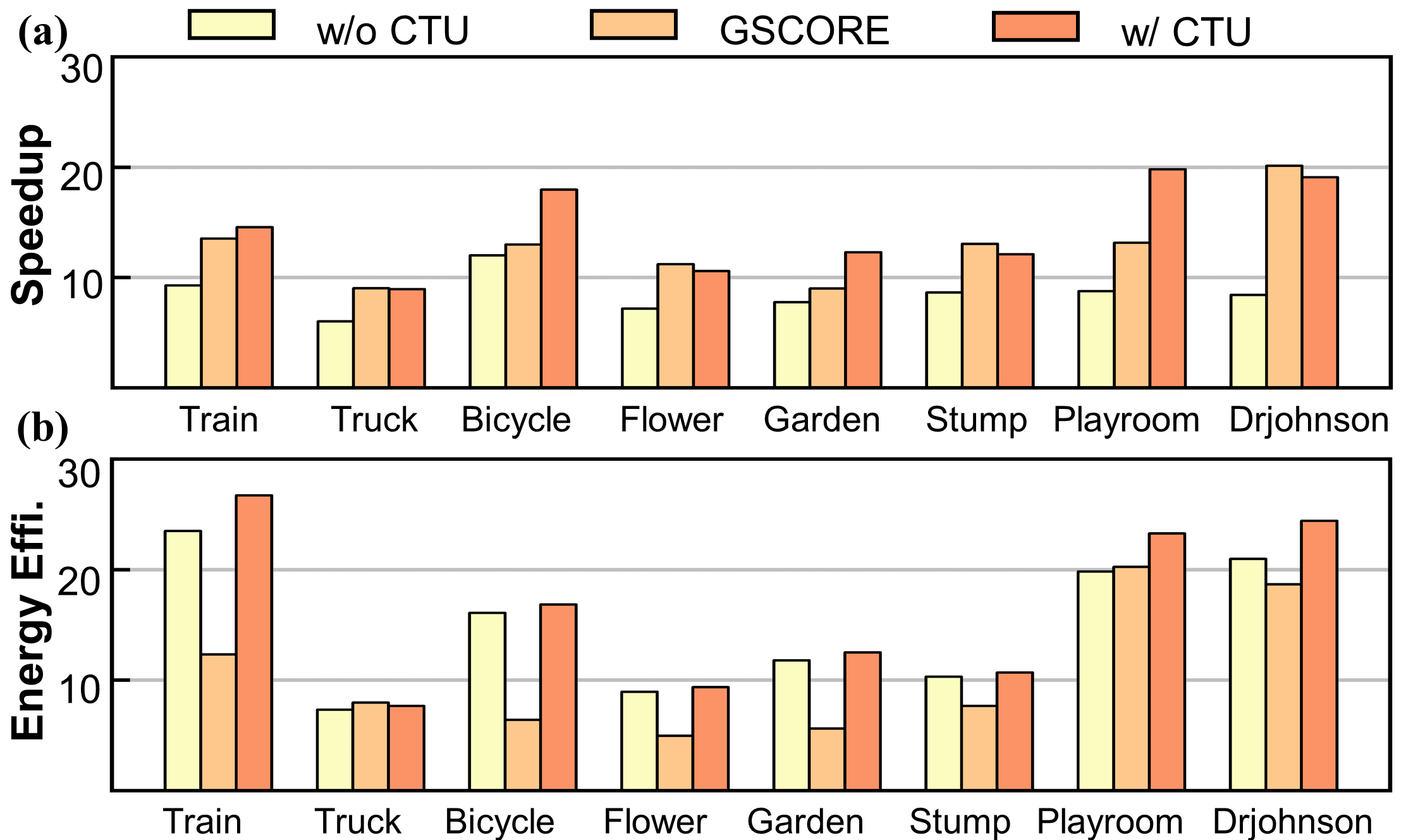}
    \caption{Overall (a) speedup and (b) energy efficiency, normalized to the GPU baseline.}
\label{Fig10_speedup_and_energy_efficiency}
\end{figure}

\begin{table}[htbp]
  \vspace{-0.3cm}
  \centering
  % 左半边：表格
  \begin{minipage}[t]{0.68\linewidth}
  \raggedright % 左对齐
  \textbf{(a)} \\
    \vspace{2mm}
    \resizebox{\linewidth}{!}{%
      \begin{tabular}{l c c}
        \Xhline{1.2pt}
        Component & Config. & Area [mm$^2$] \\ \Xhline{1.2pt}
        Preprocessing Core & 4 & 0.76 \\
        Sorting Unit & 4 & 0.16 \\
        Contri. Test Unit & 4 & 0.09 \\
        Rendering Core & 4×(4×2) & 0.96 \\
        Fea. Buffer+Others & 288KB & 1.50 \\ \Xhline{1.2pt}
        Total & & 3.47 \\ \Xhline{1.2pt}
      \end{tabular}
    }
  \end{minipage}%
  \hfill
  % 右半边：图
  \begin{minipage}[t]{0.3\linewidth} 
    \raggedright
    \textbf{(b)}
    \includegraphics[width=\linewidth]{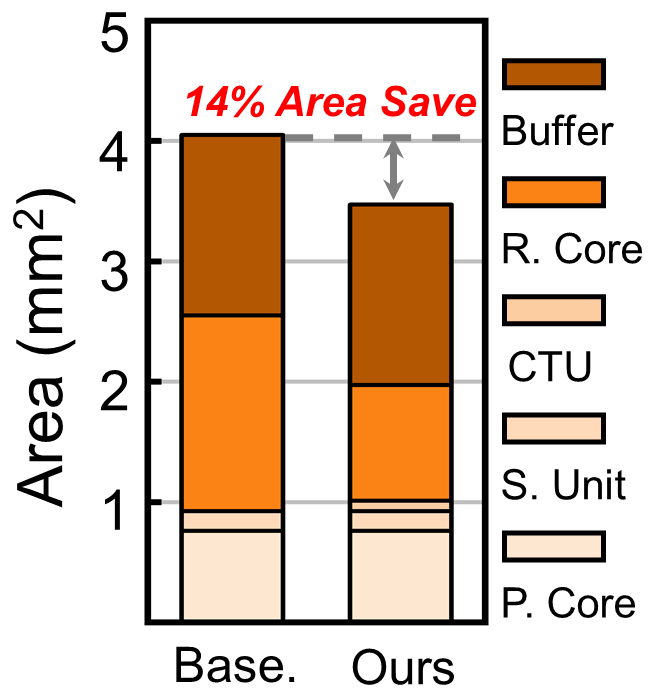}
  \end{minipage}
  \caption{(a) Hardware configuration and area breakdown. (b) Area comparison between baseline and our design.}
  \label{tab:hardware_and_area}
  \vspace{-0.5cm}
\end{table}

\subsection{Overall System Evaluation}
Tbl.~\ref{tab:psnr_ssim_comparison} compares the rendering image quality across different methods. Ours incurs only an average PSNR loss of $0.11\,\text{dB}$ compared to the pruning model, demonstrating that our adaptive leader pixel strategy is effective in capturing most contributing Gaussians and preserving visual quality.

Fig.~\ref{Fig10_speedup_and_energy_efficiency} shows the system performance over the baseline, with all values normalized to the XNX. Integrating CTU while adopting existing optimizations (pruning and clustering), FLICKER achieves average $1.1\times$ speedup over GSCore and $14.4\times$ over XNX. Furthermore, FLICKER consistently achieves the highest efficiency across all dataset, with maximum $2.6\times$ up to GSCore and $26.7\times$ compared to XNX. This demonstrates FLICKER's capability to enable real-time 3DGS rendering for edge applications, and its compatibility with existing optimization techniques.

Tbl.~\ref{tab:hardware_and_area}(a) reports the area breakdown of FLICKER. Thanks to the mixed-precision architecture and pixel-rectangle grouping, the CTU occupies less than 10\% of the VRUs area (rendering core), yet delivers up to $2.3\times$ overall speedup, which is difficult to achieve by merely adding more VRUs. We further extend the simplified version from 32 VRUs to 64 VRUs to emulate GSCore’s configuration as baseline in terms of VRU count. As shown in Tbl.~\ref{tab:hardware_and_area}(b), although the CTU and feature FIFO introduce minor additional area, they enable a more efficient area allocation than using more VRUs, ultimately achieving 14\% total area savings.

\section {Conclusion}
This paper introduces FLICKER, a contribution-aware accelerator that focuses on reducing unnecessary Gaussian processing over merely scaling parallel hardware. It performs prior contribution test to accurately skip Gaussians at fine-grained pixel blocks before rendering, while leveraging software–hardware co-design to alleviate its associated overheads. Experimental results show that our design outperforms a state-of-the-art 3DGS accelerator and an edge GPU device across most real-world datasets, while incurring less area overhead.

\newpage

\end{document}